\documentclass[preprint2,twoside]{hwo}

\usepackage{graphicx}
\usepackage{CJK}
\usepackage{bm}
\usepackage{colortbl}
\defcitealias{astro2020}{Astro2020}

\input{hwo.h}

\begin{document}

\title{\textbf{\LARGE Studying Protoplanets and Protoplanetary Disks with the \textit{Habitable Worlds Observatory}}}
\author {\textbf{\large Bin B. Ren (\begin{CJK*}{UTF8}{gbsn}任彬\end{CJK*})$^{a, b, *}$ }}
\affil{$^a$\small\it Universit\'{e} C\^{o}te d'Azur, Observatoire de la C\^{o}te d'Azur (OCA), CNRS, Laboratoire Lagrange, Bd de l'Observatoire, CS 34229, F-06304 Nice cedex 4, France; \url{bin.ren@oca.eu}}
\affil{$^b$\small\it Max-Planck-Institut f\"ur Astronomie (MPIA), K\"onigstuhl 17, D-69117 Heidelberg, Germany}
\affil{$^*$\small\it Marie Sk\l odowska-Curie Fellow}

% Please add the names of endorsers in the format "Joseph Jensen (Utah Valley University), " separated by commas.
\author{\footnotesize{\bf Endorsed by:}
Yuhiko Aoyama (Sun Yat-sen University), Jaehan Bae (University of Florida), Myriam Benisty (Max-Planck-Institut f\"ur Astronomie), Jiaqing Bi (Max-Planck-Institut f\"ur Astronomie), Aarynn Carter (Space Telescope Science Institute), Gabriele Cugno (University of Zurich), Eunjeong Lee (EisKosmos (CROASAEN), Inc.), Meredith MacGregor (The Johns Hopkins University), Ignacio Mendigut\'{i}a (Centro de Astrobiolog\'{i}a (CAB, CSIC-INTA)), Eric Nielsen (New Mexico State University), William Roberson (New Mexico State University),  Melinda Soares-Furtado (University of Wisconsin--Madison), Tomas Stolker (Leiden University), Chen Xie (The Johns Hopkins University), Shangjia Zhang (Columbia University).
}

% This section is for ADS Processing.  There must be one line per author. Leave them commented out for the present. They will be included later.
%\paperauthor{Bin B. Ren}{bin.ren@oca.eu}{0000-0003-1698-9696}{Universit\'{e} C\^{o}te d'Azur}{Observatoire de la C\^{o}te d'Azur}{Nice}{Alpes-Maritimes}{F-06304}{France}

% Please provide entries for the Author index; leave them commented out for now.
%\aindex{Ren, B. B.}

\begin{abstract}
Since the discovery of the first exoplanet orbiting a Sun-like star, the confirmation of nearly 6000 exoplanets to date -- and their diversity -- has revolutionized our knowledge of planetary systems in the past three decades. Nevertheless, the majority of these planets are around mature stars (${\gtrsim}1$~Gyr), where the planet birth environments have already dissipated. Indeed, we have only confirmed 2 forming planets (i.e., protoplanets; ${\lesssim}10$~Myr) residing in one single system. In comparison, we have imaged over 200 protoplanetary disks in the past decade, with many of them hosting substructures such as spirals and gaps which suggest the existence of protoplanets. To understand the early stages of planet formation, the \textit{Habitable Worlds Observatory} (\textit{HWO}) -- with its high-contrast imaging and integral field spectroscopy capabilities -- presents a unique opportunity to explore the demographics of the natal stages of planet formation and their birth environments. We propose to image protoplanets within substructured protoplanetary disks using \textit{HWO} via direct imaging, and characterize them (i.e., protoplanets, protoplanetary disks, circumplanetary disks) using integral field spectroscopy and spectropolarimetry. This effort will dramatically extend current population of protoplanets, probing and characterizing over 200 protoplanets. By expanding the number of protoplanets by two orders of magnitude, these observations will test and refine planet formation theory and planet-disk interaction theory, and further motivate planet migration studies together with existing mature planets. The results will offer critical insight into planetary system formation and evolution, and help understand the origin of our own Solar System. \textit{This article is an adaptation of a science case document developed for HWO's Solar System in Context (Birth \& Evolution; SSiC BE) Steering Committee.}
  \\
  \\
\end{abstract}

\vspace{2cm}

\section{Science Goal}

Despite the confirmation of nearly 6000 exoplanets,\footnote{NASA Exoplanet Archive (\url{https://exoplanetarchive.ipac.caltech.edu}), accessed 2025 June 19.\label{fn-mass-period}} only less than 40 of them are directly imaged, and we are extremely limited in understanding how they form in the earliest stage. These natal exoplanets are best accessible with \textit{HWO} by directly imaging them. We can take photos of protoplanets -- planets that are still at their formation stage -- and quantify their physical status (e.g., temperature, atmosphere), study how they shape their surroundings  (i.e., protoplanetary disks).

\subsection{Topics Related to the Astro2020:}
The topics in this Science Case are related to the 2020 Decadal Survey on Astronomy \& Astrophysics by
\citet[][hereafter \citetalias{astro2020}]{astro2020} as follows.

\begin{itemize}
    \item \textit{Planet Formation} (Section 2.1.2.1 of \citetalias{astro2020}): Understanding planet formation (e.g., how particles assemble, how the process overcomes barriers to operate quickly enough before the disk dissipates) is crucial to place potentially habitable planets in context.
    \item \textit{Connections to the Solar System} (Section 2.1.2.3 of \citetalias{astro2020}): Studying systems over a range of ages provides insights into formation mechanisms and evolutionary processes that improve the understanding of the Solar System's own history. Observations of young planet-forming disks (e.g., protoplanetary disks) provide a window into the early conditions that led to the formation of the giant planets in our Solar System.
    \item \textit{Multi-Scale Cosmic Flows of Gas} (Section 2.3.3 of \citetalias{astro2020}): A question for the coming decade is to understand the coupling between small-scale, Solar System--forming regions, the larger clound environment, and the diffuse interstellar medium (ISM).
\end{itemize}

In particular, this Science Case is relevant to the following Key Science Questions and Discovery Areas of the Astro2020 Decadal Survey Report \citepalias{astro2020}:

\begin{itemize}
    \item \textbf{E-Q1.} What Is the Range of Planetary System Architectures, and Is the Configuration of the Solar System Common?
    \begin{itemize}
        \item \textbf{E-Q1a.} What Are the Demographics of Planets Beyond the Reach of Current Surveys?
        \item \textbf{E-Q1c.} How Common Is Planetary Migration, Howe Does It Affect the Rest of the Planetary System, and What Are the Observable Signatures?
        \item \textbf{E-Q1d.} How Does the Distribution of Dust and Small Bodies in Matures Systems Connect to the Current and Past Dynamical States Within Planetary Systems?
    \end{itemize}
    \item \textbf{F-Q4.} Is Planet Formation Fast or Slow?
    \begin{itemize}
        \item \textbf{F-Q4a.} What Are the Origins and Demographics of Disk Substructures?
        \item \textbf{F-Q4b.} What Is the Range of Physical Environments Available for Planet Formation?
    \end{itemize}
    \item \textbf{Discovery Area:} Detecting and Characterizing Forming Planets
    \begin{itemize}
        \item \textbf{F-DA1.} How Do Planets and Their Satellites Grow?
        \item \textbf{F-DA2.} What Are the Atmospheres of Long-Period Giant Planets Like at Their Formation Epoch?
        \item \textbf{F-DA3.} How Do the Orbital Architectures of Planetary Systems Evolve?
    \end{itemize}
\end{itemize}

\subsection{Relevance for Other Broad Scientific Areas in the Astro2020 Decadal Survey:}
\begin{itemize}
    \item Panel on Stars, The Sun, and Stellar Populations (G)
    \item Panel on Electromagnetic Observations from Space 1 (I)
    \begin{itemize}
        \item \textbf{I.3.1.} Flagship Science Capabilities
        \item \textbf{I.3.5.} Detailed Technology Development Comments -- Starlight Suppression Considerations
    \end{itemize}
    \item Panel on Optical and Infrared Observations from the Ground (K)
    \begin{itemize}
        \item \textbf{K2.1.} Exoplanets and Astrobiology
        \item \textbf{K4.2.} Adaptive Optics/High-Contrast Imaging
    \end{itemize}
\end{itemize}

%http://localhost:8888/notebooks/Documents/2025Summer/HWO/data/sccd-fig1.ipynb
\begin{figure*}[ht!]
    \centering
    \includegraphics[width=\textwidth]{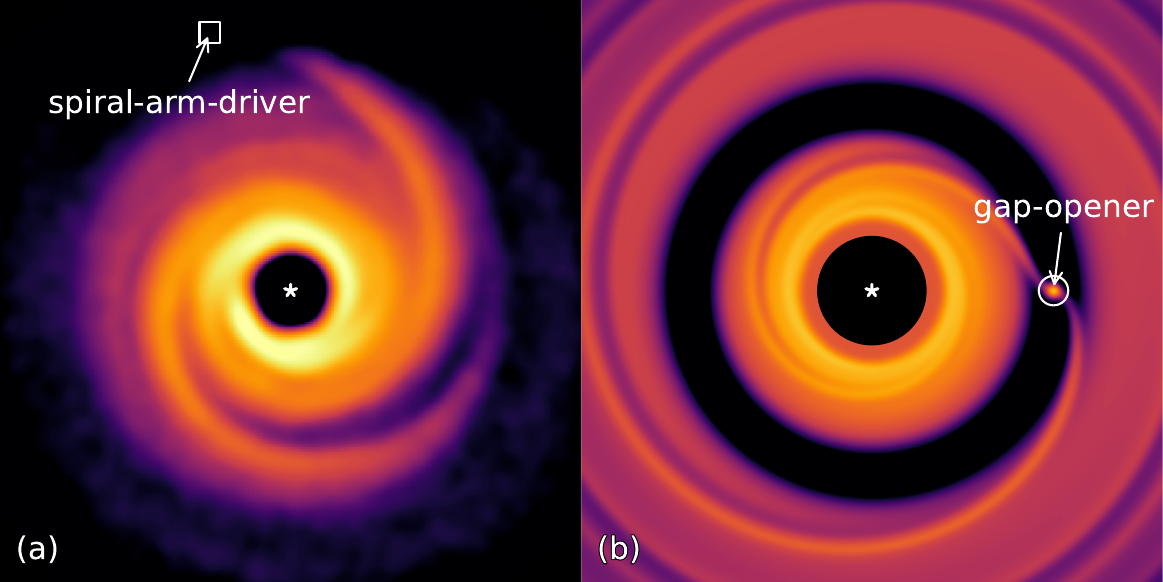}
    \caption{Substructures in protoplanetary disks, such as (a) spiral arms and (b) gaps, can inform the location and mass of protoplanets that shape these substructures \citep[marked with arrows here; e.g.,][]{dong17, bae18}. Note: the panels displayed here use simulated data provided by \citet{dong15} for (a) and \citet{bi24} for (b).}
    \label{fig:fig1}
\end{figure*}

\section{Science Objective}

Directly image protoplanets that are actively interacting and shaping their surrounding protoplanetary disks; characterize protoplanets and resolve protoplanetary disks in both spectroscopy and polarized light.

\subsection{Scarcity of Protoplanets}

Since the discovery of the first exoplanet orbiting a Sun-like star by \citet{mayor95}, astronomers have confirmed the existence of nearly 6000 exoplanets.\textsuperscript{\ref{fn-mass-period}} Despite these fruitful discoveries, the majority of discovered exoplanets are around mature stars with ages of over 1 Gyr, and only one confirmed system is less than 10 Myr \citep[the PDS~70 system with 2 young exoplanets:][]{keppler18, haffert19}, i.e., protoplanets. \textbf{The scarcity of only 2 protoplanets significantly limits our knowledge on the early stages of planet formation and planetary system evolution.}

This $6000{:}2$ exoplanet-to-protoplanet scarcity is due to the high-contrast imaging (i.e.,~direct imaging) limits from existing instruments. In spite of this scarcity, however, protoplanets are expected to be ubiquitous based on our understanding of protoplanetary disk morphology \citep[e.g.,][]{bae18}, since planets can interact with their surrounding environments and induce observable substructures in protoplanetary disks (spirals, gaps, etc. See Fig.~\ref{fig:fig1}) through planet-disk interaction.

%http://localhost:8888/notebooks/Documents/2025Summer/HWO/data/sccd-fig2.ipynb
\begin{figure*}[htb!]
    \centering
    \includegraphics[width=\textwidth]{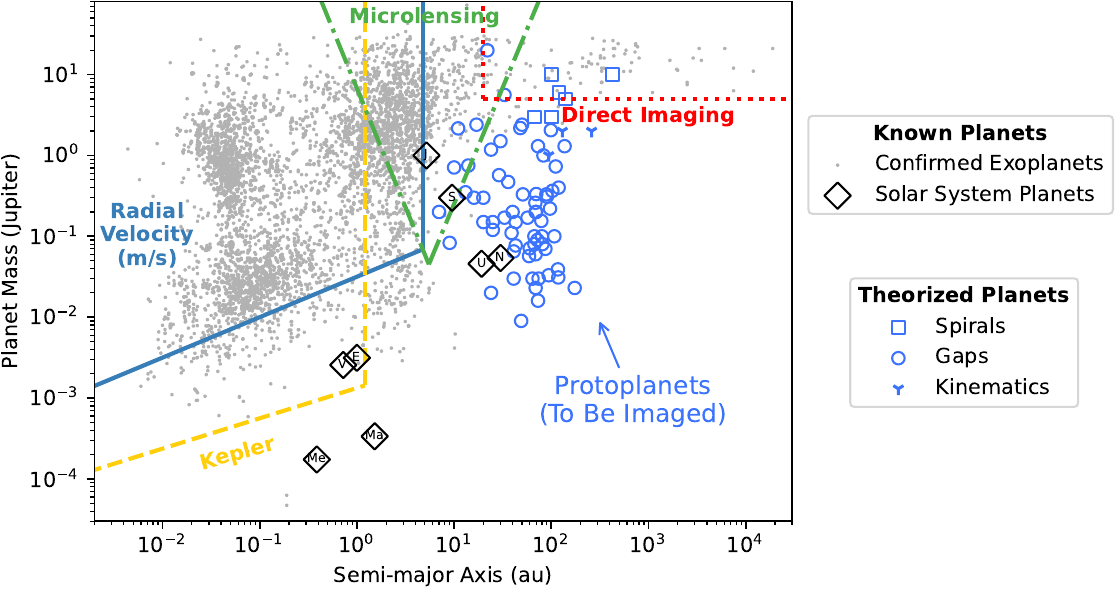}
    \caption{Protoplanets inferred from disk substructures \citep[blue symbols;][]{bae18} probe a different parameter space than confirmed exoplanets\textsuperscript{\ref{fn-mass-period}} (gray dots) and Solar System planets (black diamonds). Notes: (1) The current boundaries of different exoplanet detection methods, as well as the theorized exoplanets (non-complete sample), are provided by \citet{bae18}. (2) The \citet{bae18} figure is replotted and annotated here to specify potential protoplanets to be accessible by \textit{HWO}.}
    \label{fig:fig2}
\end{figure*}

With high-contrast, spectroscopic, and polarized light observations, \textit{HWO} can address the current scarcity of protoplanets thanks to its \textit{small inner working angle} and \textit{space-based adaptive optics} by directly imaging them. We can access the earliest stage of planet formation, probe a distinct parameter space (Fig.~\ref{fig:fig2}), and complement our knowledge of planetary system formation and evolution.

\subsection{Protoplanet and Protoplanetary Disk Imaging}
Planets can gravitationally interact with their surrounding circumstellar disks, and thus carve observable substructures which can trace the existence of exoplanets on circumstellar disks. \textit{HWO} imaging and characterization of both protoplanets and protoplanetary disks can benefit studies on exoplanets, planet-disk interaction, as well as mineral composition of circumstellar dust.

\subsubsection{Image Accreting Protoplanets}

In spite of only having confirmed and imaged two protoplanets \citep[e.g.,][]{haffert19}, many young protoplanetary disks have been imaged \citep{benisty23} that appear likely to host such unseen planets \citep[e.g.,][]{bae18}. The sensitivity of \textit{HWO} should permit us to 
\begin{itemize} 
    \item Image protoplanets in broadband;
    \item Characterize accretion signatures of protoplanets in narrow bands simultaneously (spanning from the near-ultraviolet to the near-infrared; e.g., \citealp{aoyama20}), and study their variability \citep[e.g.,][]{close25}. 
    \item Image protoplanets up to ${\sim}1000$~au to probe a distinct population from currently observed mature planets from radial velocity and transiting, and thus obtain a holistic picture of exoplanet distribution.
\end{itemize}

\subsubsection{Image Substructures in Polarized Light}
Spatially resolved protoplanetary disk imaging revealed substructures \citep[e.g.,][]{andrews20} inform the location and mass of protoplanets \citep[e.g.,][]{dong15, bae18, zhang23}. However, visible and near-infrared light imaging faces challenges in extracting high-quality data in total intensity \citep[e.g.,][]{milli12, olofsson24}. In comparison, polarized light observations 
\begin{itemize} 
    \item Provide high-quality data \citep[e.g.,][]{benisty15} to trace planetary existence;
    \item Permit dust characterization from theoretical \citep[e.g.,][]{tazaki23}, observational \citep[e.g.,][]{ren23, olofsson23}, and laboratory approaches \citep[e.g.,][]{munoz21};
    \item Separate embedded planets from disks \citep[e.g.,][]{wahhaj24}.
\end{itemize}

The combined detection and characterization of both protoplanets and protoplanetary disks with \textit{HWO} would directly benefit planet-disk interaction theory \citep[e.g.,][]{dong15, bae18}.

\subsubsection{Spectropolarimetry of Circumstellar Systems}
Spatially-resolved spectroscopy can inform mineral composition of circumstellar dust \citep[e.g., circumstellar disk photometry:][]{debes08} to relate protoplanets with their ambient environment. With additional polarimetric information \citep[e.g.,][]{chen24}, spectropolarimetry can enable a more complete understanding of 
\begin{itemize}
    \item Circumstellar/circumplanetary dust property for their mineral composition;
    \item The spatial gradient of dust properties as a function of stellocentric distance \citep[e.g.,][]{xie25}.
\end{itemize}

\textit{HWO} spectropolarimetry will offer observational evidence in bulk composition of planetary building blocks, and thus use mineralogy to inform planet migration, snow line, and planet accretion of ambient dust particles.

\begin{table*}[htb!]
    \centering
    \caption{Physical Parameters for Protoplanets and Protoplanetary disks}
    \label{tab:tab1}
\begin{tabular}{p{5.1cm}p{2.25cm}p{2.25cm}p{2.25cm}p{3.5cm}} \hline\hline
Physical Parameter                 & State of the Art                      & Incremental Progress & Substantial Progress    & \textbf{Major Progress}                     \\
               &                    & (Enhancing) & (Enabling)      & \textbf{(Breakthrough)}                     \\\hline
\rowcolor{gray!10} Number of Protoplanets             & 2                                     & 10                               & 30                                   & \textbf{200}                                               \\
Orbital and Astrometric Accuracy & $5$~au                           & $2$~au                      & $0.5$~au                        & \textbf{$\bm{0.1}$~au}                                     \\
\rowcolor{gray!10}Mass Accuracy                      & $5M_{\rm Jup}$                        & $1M_{\rm Jup}$                   & $0.1M_{\rm Jup}$                     & $\bm{0.01M_{\rm Jup}}$                                 \\
Spectral Resolution                & 30                                    & 2500                             & 5000                                 & \textbf{10\,000}                                             \\
\rowcolor{gray!10}Spectral/Accretion Features        & Selected lines (e.g., H$\alpha$ line) & Balmer series, Lyman series      & Multi-series (simultaneous)          & \textbf{Multi-epoch integral field spectroscopy}           \\
Spatial Resolution                 & $10$~au                          & $3$~au                      & $1$~au                          & \textbf{$\bm{0.1}$~au}                                     \\
\rowcolor{gray!10}Polarization Fraction Uncertainty  & $10\%$                                & $5\%$                            & $1\%$                                & \textbf{$\bm{0.1\%}$}                                           \\
Bulk Composition                   & Mass/Radius measurement               & Multi-band reflectance           & High-resolution reflectance spectrum & \textbf{High-resolution polarimetric reflectance spectrum}\\ \hline
\end{tabular}
\end{table*}

\section{Physical Parameters}

This Science Case aims to leverage \textit{HWO}'s high-contrast imaging and spectropolarimetry capabilities to detect and characterize protoplanets and protoplanetary disks, determine the spatial distribution of protoplanets, test and improve planet-disk interaction theories.

\subsection{Key Physical Parameters (Protoplanets)}
For protoplanets, the key parameters are: protoplanet number, orbital and astrometric accuracy, and accretion and spectral signatures.

\subsubsection{Protoplanet Number}
The number of protoplanets is confirmed to be 2 \citep[PDS~70:][]{haffert19, wang20}. In comparison with the over 50 inferred protoplanets that are expected from protoplanetary disk substructures \citep[e.g.,][see also Fig.~\ref{fig:fig2}]{bae18} and the nearly 6000 confirmed exoplanets, we have confirmed
\begin{itemize}
    \item (\textit{State of the Art}) 2 protoplanets
\end{itemize}
as of the writing of this Science Case in mid-2025. By accessing regions exterior to ${\sim}0.5$~au from the host stars, \textit{HWO} would reach
\begin{itemize}
    \item (\textit{Incremental}) ${\sim}10$ protoplanets with a $1 M_{\rm Jup}$ sensitivity, 
    \item (\textit{Substantial}) ${\sim}30$ protoplanets with $0.1 M_{\rm Jup}$ sensitivity, and 
    \item (\textit{Breakthrough}) ${\sim}100$ protoplanets with $0.01 M_{\rm Jup}$ sensitivity.
\end{itemize}

Given the peaking of giant exoplanet distribution at $3$--$10$~au \cite{fulton21}, which will be accessible by \textit{HWO}, as well as more recent detection and characterization of substructures in protoplanetary disks \citep[e.g.,][]{curone25, vioque25}, we will likely \textbf{image over $\bm{200}$ protoplanets -- a major breakthrough} in the field of protoplanet studies with a $0.01 M_{\rm Jup}$ ($3$ times the mass of Earth) sensitivity. Such a sensitivity can further help fully map the distribution of young planets down to $0.5$~au, supporting exoplanet migration studies with observational evidence.

\subsubsection{Orbital and Astrometric Accuracy}
Precise orbital measurement helps investigate the stability of directly imaged exoplanetary systems, and explore the potential and planning of their other observables (e.g., transiting of directly imaged planets: \citealp{wang16}).
\begin{itemize}
    \item (\textit{State of the Art}) Ground-based direct imaging observations can reach an orbital accuracy of $5$~au for detected protoplanets \citep[e.g., Keck/NIRC2:][]{wang20}.
    \item (\textit{Incremental}) With ground-based optical interferometry, the orbital accuracy is $2$~au (VLTI/GRAVITY: \citealp{wang21}).
    \item (\textit{Substantial}) The Extremely Large Telescope scheduled in ${\sim}$2030 can provide a factor of $5$ increase in astrometric accuracy in comparison with current ground-based instruments \citep{maire21}.
    \item (\textit{Breakthrough}) By reaching an accuracy of ${\leq}0.1$~au, which was only accessible with ${>}10$ years of ground-based planet imaging monitoring \citep[e.g.,][]{bonse25}, exo-Moons \citep[e.g.,][]{ruffio23} can potentially be detected by \textit{HWO} based on the astrometric change of protoplanets on daily timescales. 
\end{itemize}

\subsubsection{Protoplanet Accretion/Spectral Signatures}
\begin{figure*}[ht!]
    \centering
    \includegraphics[width=\textwidth]{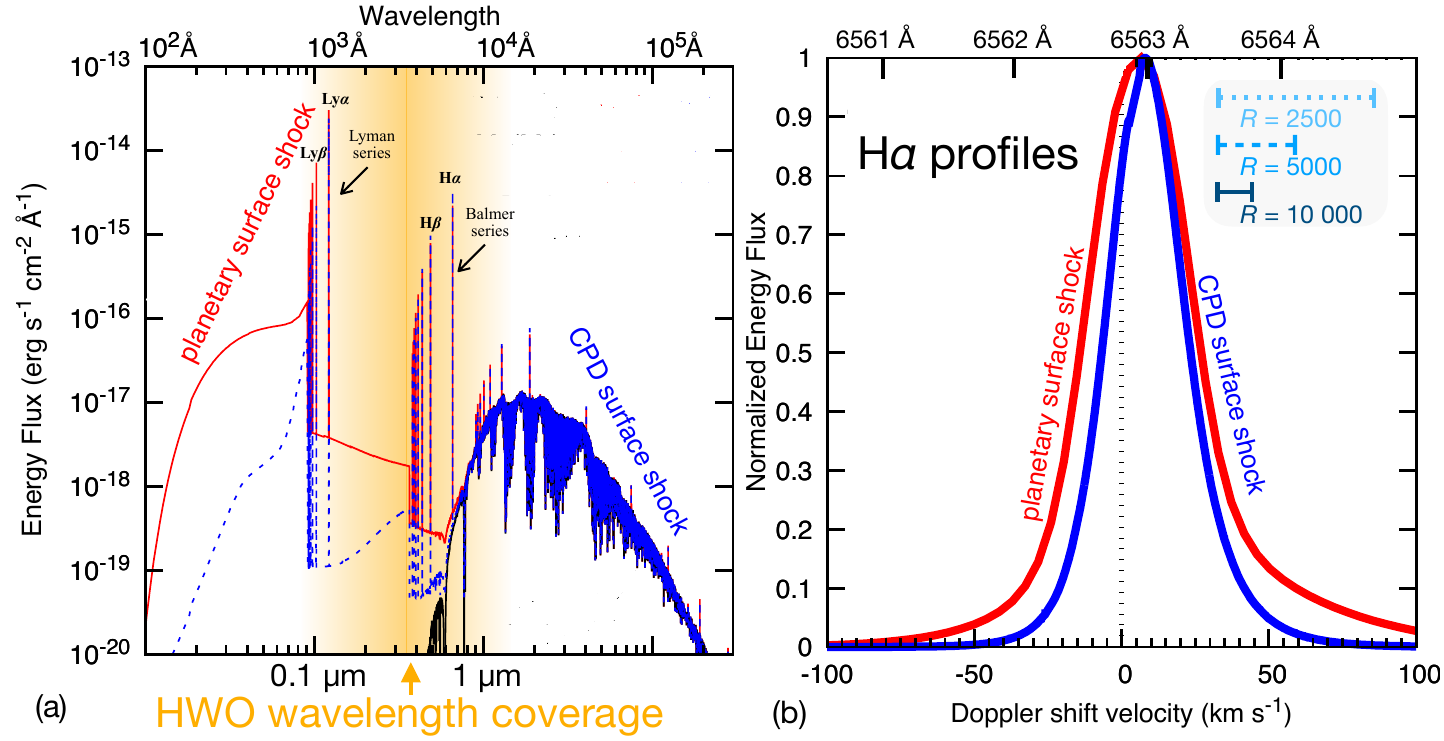}
    \caption{Two shock models (red and blue) for protoplanets resolvable by \textit{HWO} spectroscopy. (a) Spectral energy distributions or spectra. (b) H$\alpha$ profiles and different spectral resolutions: state-of-the-art VLT/MUSE instrument has $R = 2500$, which barely resolves the profile. Note: the panels here are adapted from \citet{aoyama20}.}
    \label{fig:fig3}
\end{figure*}

Spectroscopic features, including accretion features, characterize the atmosphere and accretion status for protoplanets \citep[e.g.,][]{wang21, zhou23}. For such purposes,
\begin{itemize}
    \item (\textit{State of the Art}) Low-resolution spectroscopy (spectral resolution $R\approx30$; e.g., \citealp{wang21}) in the near-infrared can explore the existence of dusty atmosphere. Narrow-band imaging tracing accretion signals (H$\alpha$; e.g., \citealp{zhou23}) constrain planetary nature.
    \item (\textit{Incremental}) Near-ultraviolet (NUV) photometry of accretion rates, together with time-series and multi-band imaging, determines accretion variability \citep[e.g.,][]{close25} and resolves the geometry of accretion flow.
    \item (\textit{Substantial}) Integral field spectroscopic observations in multiple bands can constrain accretion models with multiple tracers (Fig.~\ref{fig:fig3}a; \citealp{aoyama20}), and spectroscopically constrain models involving circumplanetary disks (CPDs).
    \item (\textit{Breakthrough}) Empirically constrain extinctions, mitigate modeling uncertainties in estimating accretion rates: this breakthrough will be reached with multi-epoch, spatially-resolved, and high-spectral resolution spectroscopy ($R \approx 10 000$; Fig.~\ref{fig:fig3}b) brings a major breakthrough in capturing spectroscopic and accretion signals. 
\end{itemize}

\subsection{Key Physical Parameters (Protoplanetary Disks)}
For protoplanetary disks, the key parameters are: protoplanet disk structure, and mineral composition.

\subsubsection{Protoplanetary Disk Substructure}
Protoplanets gravitationally interact with their forming environments, and leave observational signatures on protoplanetary disks (e.g., Fig.~\ref{fig:fig1}). High spatial resolution imaging of circumstellar disks \citep[e.g.,][]{andrews20} can trace exoplanets, helps test and improve planet-disk interaction theories.
\begin{itemize}
    \item (\textit{State of the Art}) With existing ground-based high-contrast imagers, circumstellar disks are spatially resolved at ${\sim}10$~au.
    \item (\textit{Incremental}) Tripling the pupil size of existing ground-based observatories will lead to a spatial resolution of ${\sim}3$~au.
    \item (\textit{Substantial}) At a spatial resolution of $1$~au, the leading and trailing spirals -- as well as motion -- might be resolved to test the existence of protoplanet candidates \citep[e.g.,][]{hammond23}.
    \item (\textit{Breakthrough}) With a spatial resolution of $0.1$~au, this brings a major breakthrough in resolving fine details of protoplanetary disk substructures, with a potential of resolving CPDs.
\end{itemize}

\subsubsection{Bulk Composition}
The bulk composition of exoplanets informs the mineralogy of planetary building blocks.
\begin{itemize}
    \item (\textit{State of the Art}) Existing studies used exoplanet size and mass measurements to infer bulk composition \citep[e.g.,][]{zeng19}.
    \item (\textit{Incremental}) Multi-band reflectance and existing reflectance spectrum rules out certain compositions for solid materials in circumstellar systems \citep[e.g.,][]{debes08}.
    \item (\textit{Substantial}) High-resolution reflectance spectrum in total intensity helps characterize materials.
    \item (\textit{Breakthrough}) High-resolution polarimetric and total intensity spectrum (i.e., spectropolarimetry) brings a major breakthrough in characterizing dust particles in circumstellar systems.
\end{itemize}

The physical parameters for both protoplanets and protoplanetary disks are summarized in Table~\ref{tab:tab1}.

\section{Description of Observations}

With \textit{HWO}, we can reach breakthrough science in the protoplanet and protoplanetary system imaging and integral field spectroscopy studies with large field of view, polarimetry (and spectropolarimetry), high spatial resolution, high spectral resolution (into the near-ultraviolet), small inner working angle, large outer working angle, etc. With space-based adaptive optics, we should target young stars with $V$-band magnitude $\lessapprox 15$.

\subsubsection*{Large Field of View (imaging)}
The field of view (FoV) for high-contrast imagers are
\begin{itemize}
    \item (\textit{State of the Art}) $12\farcs5\times12\farcs5$, from Very Large Telescope (VLT)/SPHERE's IRDIS instrument (1024 pixels, with 1 pixel = 12.25 mas). Future imaging FoV should reach
    \item (\textit{Incremental}) $0\farcm5\times0\farcm5$, to capture the largest known angular extent of a protoplanetary disk -- AB Aur -- and image protoplanets within.
    \item (\textit{Substantial}) $1'\times1'$, to match the FoV of the \textit{Hubble Space Telescope} (\textit{HST})/STIS coronagraph.
    \item (\textit{Breakthrough}) $2'\times2'$, to ensure the capturing of the largest known angular extent of a circumstellar disk -- Fomalhaut.
\end{itemize}

\begin{table*}[htb!]
    \centering
    \caption{Observation Requirements for Protoplanets and Protoplanetary disks}
    \label{tab:tab2}
\begin{tabular}{p{4.5cm}p{2.1cm}p{2.1cm}p{2.1cm}p{2.1cm}} \hline\hline
Observation Requirement                 & State of the Art                      & Incremental Progress & Substantial Progress    & \textbf{Major Progress}                     \\
               &                    & (Enhancing) & (Enabling)      & \textbf{(Breakthrough)}                     \\\hline
\rowcolor{gray!10} Large field of view (imaging)             & $12\farcs5\times12\farcs5$                                     & $0\farcm5\times0\farcm5$                               & $1\arcmin\times1\arcmin$                                   & $\bm{2\arcmin\times2\arcmin}$                                               \\
Large field of view (integral field spectrograph) & $2\arcsec\times2\arcsec$                           & $5\arcsec\times5\arcsec$                     & $10\arcsec\times10\arcsec$                        & $\bm{2\arcmin\times2\arcmin}$                                     \\
\rowcolor{gray!10}Polarimetric mode                      & \multicolumn{2}{c}{Polarimetry}                   & \multicolumn{2}{c}{\textbf{Spectropolarimetry}}                                 \\
Spatial resolution (at $500$~nm)                & $50$~mas                                    & $20$~mas                             & $10$~mas                                 & \textbf{$\bm{5}$~mas}                                             \\
\rowcolor{gray!10}Spectral feature observation        & \multicolumn{2}{c}{Narrow-band and multi-band}      & Multi-band (simultaneous)          & \textbf{Spectroscopy}           \\
Spatial Resolution                 & $10$~au                          & $3$~au                      & $1$~au                          & \textbf{$\bm{0.1}$~au}                                     \\
\rowcolor{gray!10}Spectral resolution (with AO)  & $30$                                & $100$                            & $3000$                                & \textbf{$\bm{10\,000}$}                                           \\
Inner working angle       & $0\farcs15$               & $0\farcs05$          & $0\farcs02$ & $\bm{0\farcs005}$\\
\rowcolor{gray!10} Outer working angle       & $1\arcsec$               & $2\arcsec$          & $5\arcsec$ & $\bm{1\arcmin}$\\
$V$-mag of host star      & $5$               & $8$          & $12$ & $\bm{15}$\\ \hline
\end{tabular}
\end{table*}

\subsubsection*{Large Field of View (integral field spectrograph, IFS)}
The FoVs for high-contrast IFS imagers are
\begin{itemize}
    \item (\textit{State of the Art}) $2\arcsec\times2\arcsec$, from VLT/SPHERE's IFS instrument (290 pixels, with 1 pixel = 7.46 mas). Future IFS FoVs should reach
    \item (\textit{Incremental}) $5\arcsec\times5\arcsec$, to capture the majority of the brightest protoplanetary disks in scattered light \citep[e.g.,][]{ren23}.
    \item (\textit{Substantial}) $10\arcsec\times10\arcsec$, to match the largest angular extent of the AB Aur protoplanetary disk from SPHERE to spectroscopically characterize protoplanets.
    \item (\textit{Breakthrough}) $1'\times1'$, to ensure the capturing of protoplanets in the AB Aur system, seen with \textit{HST}/STIS.
\end{itemize}

\subsubsection*{Polarimetric Mode}
The polarimetric imaging mode of high-contrast imagers in 2015 -- 2025 has revolutionized the field of protoplanetary system imaging \citep[e.g.,][]{benisty15}, the modes are
\begin{itemize}
    \item (\textit{State of the Art} and \textit{Incremental}) polarimetric imaging, and we should aim towards
    \item (\textit{Substantial} and \textit{Breakthrough}) spectropolarimetric imaging to combine IFS with polarimetry and characterize protoplanetary systems. 
\end{itemize}
We can reach a \textbf{major breakthrough with spectropolarimetry} in separating different polarization signals from protoplanets and their surrounding protoplanetary disks. This is essential since protoplanets are expected to be embedded in protoplanetary disks (e.g., Fig.~\ref{fig:fig1}b), and without such a proper separation, certain protoplanets have been previous missed (e.g., PDS~70~c was not recovered in \citealp{keppler18} due to its embedding).

\subsubsection*{Spatial Resolution (at ${\sim}500$~nm)}
The spatial resolution for a space-based high-contrast imager is
\begin{itemize}
    \item (\textit{State of the Art}) $50$~mas, from the pixel size of \textit{HST}/STIS coronagraph for a pupil with $2.4$~m diameter. Future spatial resolution should reach
    \item (\textit{Incremental}) $20$~mas, for a Nyquist-sampled $2.4$~m pupil.
    \item (\textit{Substantial}) $10$~mas, for a ${\sim}5$~m pupil with Nyquist sampling.
    \item (\textit{Breakthrough}) $5$~mas, for a ${\sim}10$~m pupil with Nyquist sampling.  While a $10$~m pupil is beyond the design for \textit{HWO}, this resolution can be reached at a wavelength of ${\sim}200$~nm.
\end{itemize}

\subsubsection*{Spectral Signature}
The accretion signature and status of protoplanets can be characterized with narrow-band imaging or spectroscopy.
\begin{itemize}
    \item (\textit{State of the Art} and \textit{Incremental}) Narrow-band H$\alpha$ line imaging (\textit{HST}/WFC3's F656N channel: $656$~nm $\pm2$~nm), as well as other accretion bands \citep{zhou23}. Incremental progress includes multiple accretion bands, but not simultaneously observed.
    \item (\textit{Substantial}) Multi-band observations at the same time.
    \item (\textit{Breakthrough}) Spectroscopic observations covering Lyman and Balmer series.
\end{itemize}

\subsubsection*{Spectral Resolution (Equipped with Adaptive Optics, AO)}
The spectral resolution of AO-equipped instruments are currently
\begin{itemize}
    \item (\textit{State of the Art}) the IFS instruments on VLT/SPHERE and Gemini Planet Imager (GPI), with spectral resolution $R \approx 30$.
    \item (\textit{Incremental}) $R \approx 100$ IFS (e.g., \textit{JWST}/NIRSpec IFU, which is not equipped with AO) allows reflectance spectrum characterization of circumstellar environments.
    \item (\textit{Substantial}) $R \approx 3000$ IFS (e.g., VLT/MUSE, not equipped with AO) allows accretion signature detection, see Fig.~\ref{fig:fig3}b.
    \item (\textit{Breakthrough}) $R \approx 10\,000$ IFS brings a major breakthrough in characterizing the spectral features for accretion lines, see Fig.~\ref{fig:fig3}b.
\end{itemize}

\subsubsection*{Inner Working Angle (IWA)}
The IWA defines the closest-in region that a coronagraph can access.
\begin{itemize}
    \item (\textit{State of the Art}) IWA $= 0\farcs15$ is a typical IWA for VLT/SPHERE and GPI.
    \item (\textit{Incremental}) IWA $= 0\farcs05$ is an incremental change for $1$~au regions to access terrestrial planets for stars within $20$~pc.
    \item (\textit{Substantial}) IWA $= 0\farcs02$ has substantial improvement for $3$~au regions (Solar System giant planets) for star-forming regions at ${\sim}140$~pc.
    \item (\textit{Breakthrough}) IWA $= 0\farcs005$ brings a breakthrough by accessing the $1$~au region for terrestrial protoplanets for star-forming regions at ${\sim}140$~pc.
\end{itemize}

\subsubsection*{Outer Working Angle (OWA)}
The OWA defines the furthest-out region that an AO system can control the speckles.
\begin{itemize}
    \item (\textit{State of the Art}) OWA $= 1\arcsec$ is the $H$-band OWA for the VLT/SPHERE IRDIS instrument.
    \item (\textit{Incremental}) OWA $= 2\arcsec$ accesses protoplanets up to ${\sim}300$~au regions for star-forming regions at ${\sim}140$~pc.
    \item (\textit{Substantial}) OWA $= 5\arcsec$ brings substantial improvement to explore ${\sim}500$~au regions for star-forming regions at ${\sim}140$~pc.
    \item (\textit{Breakthrough}) OWA $= 1\arcmin$ brings a breakthrough by accessing the largest angular extent of circumstellar disks that can image potential protoplanets.
\end{itemize}

\subsubsection*{Magnitude of Host Star (Space-based AO)}
There are no space-based AO systems as of the time of this Science Case.
\begin{itemize}
    \item (\textit{State of the Art}) Before \textit{HWO} launch, the Coronagraph Instrument (CGI) on \textit{Roman Space Telescope} in ${\sim}2027$ is limited to an apparent magnitude of stellar $V \lesssim 5$ \citep{krist23}.
    \item (\textit{Incremental}) $V \approx 8$ can reach the ground-based extreme AO operation ranges from SPHERE and GPI.
    \item (\textit{Substantial}) $V \approx 12$ remarks substantial increment to access the current extreme  AO limit from the ground.
    \item (\textit{Breakthrough}) $V \approx 15$ reaches the majority of the faint stars in the star-forming regions at ${\sim}140$~pc.
\end{itemize}

The observational requirements for both protoplanets and protoplanetary disks are summarized in Table~\ref{tab:tab2}.

\subsection{Why HWO Is Needed}

Being a space-based observatory equipped with adaptive optics system, \textit{HWO}'s high contrast, high spectral resolution, and high spatial resolution imaging and spectroscopy make it the only facility in providing high quality datasets for the detection and characterization of protoplanets and protoplanetary disks:

\begin{itemize}
    \item Equipped with space-based adaptive optics, \textit{HWO} offers stable point spread functions for exoplanetary system imaging, especially for the faintest and most populous host stars.
    \item Due to atmospheric absorption for ground-based observatories, the near-ultraviolet observations for protoplanet accretion features can only be executed from the space.
\end{itemize}

%\acknowledgements
{\bf Acknowledgements.} We thank comments and suggestions (including but not limited to accretion, spectroscopy, polarimetry, outer working angle) from Yifan Zhou, Chen Xie, Gregory Herczeg, Yasuhiro Hasegawa, Karl Stapelfeldt, Kate Follette, and Courteney Dressing that improved the original \textit{HWO} Science Case Development Document (SCDD; \#SCDD-SSiC-8). See more SCDDs at \url{https://outerspace.stsci.edu/display/HWOCOMMUNITYSCI/HWO+Community+Science+Case+Portal}. B.B.R.~thanks Yuhiko Aoyama, Jiaqing Bi, and Ruobing Dong for providing datasets and their permission in replotting their published figures here in Figs.~\ref{fig:fig1}--\ref{fig:fig3}. This research has received funding from the European Union's Horizon Europe research and innovation programme under the Marie Sk\l odowska-Curie grant agreement No.~101103114.

\bibliography{refs}

\end{document}